\documentclass[amsmath,amssymb]{PoS}
\usepackage{bm}
\usepackage{bbm}     
\usepackage{amsthm}
\usepackage{mathrsfs}

\title{Beyond QCD: A Composite Universe}

\ShortTitle{Beyond QCD: A Composite Universe}

\author{\speaker{Francesco Sannino}\\
    CP$^3$-Origins and the Danish Institute for Advanced Study, DIAS, University of Southern Denmark, Campusvej 55, 5230 Odense M, Denmark\\
        E-mail: \email{sannino@cp3.dias.sdu.dk}} 

\abstract{Strong dynamics constitutes one of the pillars of the standard model of particle interactions, and it accounts for the bulk of the visible matter in the universe. It is therefore a well posed question to ask if the rest of the universe can be described in terms of new highly natural four-dimensional strongly coupled theories. The goal of this brief summary is to provide, for the first time, a coherent overview of how new strong dynamics can be employed to address the relevant challenges in particle physics and cosmology from composite Higgs dynamics to dark matter and inflation. 
~
 \vskip .2cm {\small \it  Prerint numbers: CP3-Origins-2012-01 \& DIAS-2012-01.}} 

\FullConference{International Workshop on QCD Green's Functions, Confinement and Phenomenology,\\
		September 05-09, 2011\\
		Trento Italy}

\begin{document}

\section{Quantum Chromodynamics}
Quarks and gluons constitute the building blocks of ordinary matter and their interaction is governed by Quantum Chromodynamics (QCD). QCD is precisely a Yang-Mills theory with fermionic Dirac matter, the quarks, transforming according to the fundamental representation of the  $SU(3)$ color gauge group.  At low energies the theory is strongly coupled while is weak at high energies. This phenomenon is known as asymptotic freedom \cite{Gross:1973id,Politzer:1973fx}. The knowledge of the perturbative regime of the theory is relevant, for example, to disentangle at the Large Hadron Collider (LHC) experiments the proper standard model background from signals of new physics. However, it is almost always the nonperturbative regime the one which has fascinated generation of physicists while still remaining unconquered. The Nambu-Jona-Lasinio model, dual models, string theory, strong coupling expansion, lattice field theory, non-relativistic quark model, MIT bag-model, large number of colors limit, heavy quark limits, effective Lagrangians, Skyrmions, 't Hooft anomaly matching conditions, Dyson-Schwinger approximation, and more recently the holographic approach have been devised to tackle different non-perturbative aspects of this fundamental theory of nature. All these methods have limitations being either rough approximations or, if precise, at best able to explore certain dynamical or kinematical regimes of the theory. It is for this reason that some of the most hunted properties of the theory remain still unexplained. To mention a few: Is there a relation between confinement\footnote{Technically QCD does not confine, meaning, that the {\it would be} order parameter, the Polyakov loop, is not a true order parameter for confinement \cite{Sannino:2002wb}. This happens because the quarks transform according to the fundamental representation of the color gauge group.  Confinement for this system has to be understood more like the vestige of the true confinement mechanism operating for the pure Yang-Mills theory, i.e. the $SU(3)$ Yang-Mills theory without quarks.} and chiral symmetry breaking \cite{Sannino:2002wb,Mocsy:2003qw} ?; What happens when we squeeze or heat up strongly interacting matter \cite{Alford:2007xm}? Are there {\it magnetic}-like descriptions in terms of alternative weakly coupled theories \cite{Sannino:2009qc,Sannino:2010fh,Komargodski:2010mc,Sannino:2011mr,Mojaza:2011rw}? Are there higher dimensional {\it gravitational} dual descriptions of strong interactions? These are fundamental questions, however, as I shall detail in the next section there are a number of relevant areas of research craving for a broader understanding of strongly interacting theories of non-supersymmetric nature.

\section{Why the need to go beyond QCD?  }
Circa 96\% of the universe is made by unknown forms of matter and energy, while to describe the remaining 4\% one needs at least three fundamental forces, i.e. Quantum Electrodynamics (QED), Weak Interactions and QCD. Furthermore strong interactions are responsible for creating the bulk of the bright mass, i.e. the 4\%. It is therefore natural to expect that to correctly describe the rest of our universe while providing a sensible link to the visible component new forces will soon emerge.  There are, in my view, at least three primary areas of research where new strong dynamics can be efficiently deployed. The first is the sector responsible for breaking spontaneously the electroweak symmetry. The standard model Higgs in this scenario is expected to be replaced by new strongly interacting dynamics. The second application is in the use of strongly interacting dynamics to construct (near) stable dark matter candidates. Last but not the least I envision the possibility that even the mechanism behind inflation emerges from new strong dynamics.

\subsection{Composite Higgs}

Does the electroweak symmetry break in the complete absence of the standard model Higgs sector? The answer is {\it Yes}.  QCD already breaks the electroweak symmetry spontaneously as we shall momentarily show. To better elucidate this fact let's start from the symmetries of the standard model Higgs when the gauge couplings are switched off. The symmetry group is $SU(2)_L\times SU(2)_R$ which can be made explicit when re-writing the Higgs
doublet field \begin{eqnarray} H=\frac{1}{\sqrt{2}}\left(%
\begin{array}{c}
  \pi_2 + i\, \pi_1 \\
  \sigma - i\, \pi_3 \\
\end{array}%
\right)\end{eqnarray} as the right column of the following two by two matrix:
\begin{eqnarray}
\frac{1}{\sqrt{2}}\left(\sigma + i\,
\vec{\tau}\cdot\vec{\pi} \right) \equiv M
 \ .
\end{eqnarray}
The first column can be identified with the column vector $\tau_2H^{\ast}$ while the second with $H$.   $\tau^2$ is the second Pauli matrix. 
The $SU(2)_L\times SU(2)_R$ group acts linearly on $M$ according
to
\begin{eqnarray}
M\rightarrow g_L M g_R^{\dagger} \qquad {\rm and} \qquad g_{L/R} \in SU_{L/R}(2)\ .
\end{eqnarray}
One can verify that:
\begin{eqnarray}
M\frac{\left(1-\tau^3\right)}{2} = \left[0\ , \, H\right] \ . \qquad
M\frac{\left(1+\tau^3\right)}{2} = \left[i\,\tau_2H^{\ast} \ , \, 0\right] \ .
\end{eqnarray}
In QCD with two light quarks (up and down) the quantum global symmetry group is exactly $SU(2)_L\times SU(2)_R$ up to the baryon number $U(1)_B$. A similar matrix $M$ can be constructed using the quarks bilinears which, in a suggestive form, reads 
\begin{equation}
\sigma_{QCD} \sim \bar{q}q \ , \qquad \qquad \vec{\pi}_{QCD} \sim i\, \bar{q}\vec{\tau} \gamma_5 q \ ,
 \end{equation}
building the QCD meson matrix $M_{QCD}$. This symmetry is known as the chiral symmetry of QCD which experiments have first indicated to break spontaneously to an unbroken subgroup $SU(2)_V$. In this case, when setting to zero the up and down quark masses, there are three Goldstone modes ($\vec{\pi}$).
  
 Turning  on the electroweak interactions is done by simply recalling that the quarks up and down form a weak doublet. Therefore the $SU(2)_L$ symmetry is gauged by introducing the weak gauge
bosons $W^a$ with $a=1,2,3$. The hypercharge generator is taken to
be the third generator of $SU(2)_R$. The ordinary covariant
derivative acting on the standard model Higgs (M) acts identically on the QCD matrix $M_{QCD}$ and we have:
\begin{eqnarray}
D_{\mu}M_{QCD}=\partial_{\mu}M_{QCD} -i\,g\,W_{\mu}M_{QCD} + i\,g^{\prime}M_{QCD}\,B_{\mu} \ , \qquad {\rm
with}\qquad W_{\mu}=W_{\mu}^{a}\frac{\tau^{a}}{2} \ ,\quad
B_{\mu}=B_{\mu}\frac{\tau^{3}}{2} \ . \nonumber \\
\end{eqnarray}
 
 Recalling the fact that chiral symmetry is dynamically broken in QCD we have:
\begin{equation}
\langle \sigma_{QCD} \rangle^2 \equiv v_{QCD}^2 \qquad {\rm and} \qquad \sigma_{QCD}= v_{QCD} + h_{QCD} \ ,
\end{equation} 
where $h_{QCD}$ is a QCD scalar meson. It is a bilinear, in the quarks, excitation around the vacuum expectation value of $\sigma_{QCD}$ and remains heavy\footnote{The nature of this scalar state constitutes a phenomenological important puzzle in QCD. This is so since this state possesses the quantum number of the vacuum and therefore mixes with several other states of the theory made of higher quark and glue Fock states \cite{Sannino:1995ik,Harada:1995dc,Harada:2003em,Sannino:2007yp}. However there is a simple and clear limit in which some light can be shed. This is the 't Hooft mathematical large number of colors limit. In this limit all the mesons made by quark bilinears become stable and non-interacting. Therefore we expect this state to exist also when reducing the number of colors to three. One possibility is to identify it with $f_0(1300)$. Of course this state at any finite number of colors will never be a pure bilinear one.  There is much confusion in the literature about this quark bilinear which is often erroneously identified with the $f_0(600)$ resonance. This state is not (mostly) a quark bilinear but its nature and physical properties are closer to the ones expected from a multi-quark state nature\cite{Sannino:1995ik,Harada:1995dc,Harada:2003em,Sannino:2007yp}.  $f_0(600)$ is, however, a crucial ingredient in the unitarization of pion-pion scattering at low energies and this fact is not directly related to the composition of the scalar meson but simply with its overall quantum numbers \cite{Sannino:1995ik,Harada:1995dc,Harada:2003em,Sannino:2007yp} under space-time symmetries and the unbroken $SU(2)_V$ global symmetry. }.

In this way we have achieved that the global symmetry breaks dynamically to its diagonal subgroup:
\begin{eqnarray}
SU(2)_L\times SU(2)_R \rightarrow SU_V(2) \ .
\end{eqnarray}
To be more precise the $SU(2)_R$ symmetry is already broken explicitly by our choice of gauging only an $U_Y(1)$ subgroup of it and hence the actual symmetry breaking pattern is:
\begin{eqnarray}
SU(2)_L\times U_Y(1) \rightarrow U_Q(1) \ ,
\end{eqnarray}
with $U_Q(1)$ the electromagnetic abelian gauge symmetry. According to the Nambu-Goldstone's theorem three massless degrees of freedom appear, i.e. $\pi^{\pm}_{QCD}$ and $\pi^0_{QCD}$. In the unitary gauge these Goldstones become the longitudinal degree of freedom of the massive elecetroweak gauge-bosons. Substituting the vacuum value for $\sigma_{QCD}$ in the Higgs Lagrangian the gauge-bosons quadratic terms read:
\begin{equation}
\frac{v_{QCD}^2}{8}\, \left[g^2\,\left(W_{\mu}^1
W^{\mu,1} +W_{\mu}^2 W^{\mu,2}\right)+ \left(g\,W_{\mu}^3 -
g^{\prime}\,B_{\mu}\right)^2\right]  \ . \end{equation}
 The $Z_{\mu}$ and the photon $A_{\mu}$ gauge bosons are:
\begin{eqnarray}
Z_{\mu}&=&\cos\theta_W\, W_{\mu}^3 - \sin\theta_{W}B_{\mu} \ ,\nonumber \\
A_{\mu}&=&\cos\theta_W\, B_{\mu} + \sin\theta_{W}W_{\mu}^3 \ ,
\end{eqnarray}
with $\tan\theta_{W}=g^{\prime}/g$ while the charged massive vector bosons are
$W^{\pm}_{\mu}=(W^1\pm i\,W^2_{\mu})/\sqrt{2}$. 
The bosons masses $M^2_W=g^2\,v_{QCD}^2/4$ due to
the custodial symmetry satisfy the tree level relation $M^2_Z=M^2_W/\cos^2\theta_{W}$.  

What is missing to have a phenomenologically successful explanation of the observed spontaneous breaking of the electroweak gauge symmetry?: 

\begin{itemize}

\item The scale of the QCD condensate is too small to be able to account for the observed mass of the electroweak gauge bosons.

\item We observed experimentally three light (pseudo) Goldstone bosons identified with the QCD pions. 

\item The quarks themselves have masses which means that another mechanism is in place for giving masses to the standard model fermions. 

\end{itemize}
The first two issues can be resolved by postulating the existence of yet another strongly coupled theory with a dynamical scale taken to be the electroweak one ($v_{QCD} \rightarrow v_{weak}$), while the third one requires yet another sector. Weinberg \cite{Weinberg:1979bn} and Susskind  \cite{Susskind:1978ms} considered a new copy of QCD by simply rescaling the invariant mass of the theory, e.g. the new proton mass, and dubbed the model Technicolor. However this model is at odds with experiments as summarized in \cite{Sannino:2009za} while the idea of dynamical breaking of the electroweak theory is very much alive. 
Whatever the new sector responsible for dynamical electroweak symmetry breaking is it will mix with the QCD one. In fact, the observed QCD physical pions are linear combination of the QCD pion eigenstates and the ones emerging from the new sector. The mixing is of the order of $v_{QCD}/v_{weak} \sim 10^{-3}$. Through this mixing the physical pion wave function, if measured with great accuracy, should be able to reveal the mechanism behind electroweak symmetry breaking using low energy data. 

If the new gauge dynamics contains only fermionic matter than quantum corrections lead, at most, to logarithmic corrections to the bare parameters of the theory, and therefore the theory is said to be technically natural. A layman version of this statement is that: {\it small parameters stay small under renormalization}. 

Confronting models of dynamical electroweak symmetry breaking with precision measurements as well as the challenge of generating the standard model fermion masses without generating, in the quark sector, large flavor changing neutral currents has led to defining the framework of {\it ideal walking} \cite{Fukano:2010zz}. According to this paradigm a reasonable consistent scenario for models of dynamical electroweak symmetry breaking without large fine-tuning requires the new strongly coupled theory, in isolation, to display large distance conformality while the sector needed to give masses to the standard model fermions will provide the operators needed to deform away from conformality the new strongly coupled dynamics. The original idea of {\it walking} \cite{Holdom:1981rm} suffers from a number of shortcomings summarized in \cite{Fukano:2010zz}. By marrying minimal models of technicolor \cite{Sannino:2004qp,Hong:2004td,Dietrich:2005jn,Foadi:2007ue}  (i.e. the ones featuring the smallest conformal S-parameter \cite{Sannino:2010ca,Sannino:2010fh,DiChiara:2010xb,Sondergaard:2011ps}) with supersymmetry at the scale of fermion mass generation in  \cite{Antola:2011bx,Antola:2010nt}  the first explicit example of ultraviolet complete theory of (ideal) walking type  has been constructed. 

\subsection{Composite Dark Matter}

Experimental observations strongly indicate that  the  universe is flat and predominantly made of unknown forms of matter. Defining with $\Omega$ the ratio of the density to the critical density, observations indicate that the fraction of matter amounts to  $\Omega_{\rm matter} \sim 0.3$ of which the normal baryonic one is only $\Omega_{\rm baryonic} \sim 0.044$. The amount of non-baryonic matter is termed dark matter. The total $\Omega$  in the universe is dominated by dark matter and pure energy (dark energy) with the latter giving a contribution $\Omega_{\Lambda} \sim 0.7$ (see for example \cite{Frieman:2008sn,Frieman:2002wi}). Most of the  dark matter is ``cold''  (i.e. non-relativistic at freeze-out) and significant fractions of hot dark matter are hardly compatible with data. What constitutes dark matter is a question relevant for particle physics and cosmology. A WIMP (Weakly Interacting Massive Particles) can be the dominant part of the non-baryonic dark matter contribution to the total $\Omega$. Axions can also be dark matter candidates but only if their mass and couplings are tuned for this purpose.

It would be theoretically very pleasing to be able to relate
the dark matter and the baryon energy densities in order to  explain the ratio
$\Omega_{\rm DM}/\Omega_B\sim 5$~\cite{Komatsu:2008hk}. We know that the amount of baryons in the universe $\Omega_B\sim 0.04$ is
determined solely by the cosmic baryon asymmetry $n_B/n_\gamma \sim 6\times
10^{-10}$.  This is so since the baryon-antibaryon annihilation cross section
is so large, that virtually all antibaryons annihilate away, and only the
contribution proportional to the asymmetry remains.  
This asymmetry can be
dynamically generated after inflation.  We do not know, however, if the dark matter density is determined by thermal freeze-out,
by an asymmetry, or by something else.
Thermal freeze-out needs a $\sigma v \approx 3~10^{-26}\, {\rm cm}^3/{\rm sec}$ which is of the electroweak scale,
suggesting a dark matter mass in the TeV range. If $\Omega_{\rm DM}$ is determined by thermal freeze-out, its proximity to $\Omega_B$
is just a fortuitous coincidence and is left unexplained.

If instead $\Omega_{\rm DM}\sim \Omega_B$ is not accidental, then the theoretical
challenge is to define a consistent scenario in which the two energy densities
are related.  Since $\Omega_B$ is a result of an asymmetry, then relating the
amount of dark matter to the amount of baryon matter can very well imply that
$\Omega_{\rm DM} $ is related to the same asymmetry that determines $\Omega_B$.
Such a condition is straightforwardly realized if the asymmetry for the dark matter
particles is fed in by the non-perturbative electroweak sphaleron transitions,
that at temperatures much larger than the temperature $T_*$ of the electroweak
phase transition (EWPT) equilibrates the baryon, lepton and dark matter
asymmetries.  Implementing this condition implies the following requirements: 
\begin{itemize}
\item[1.] dark matter must be (or must be a composite state of) a fermion, 
chiral (and thereby non-singlet) under the weak $SU(2)_{L}$, 
and carrying an anomalous (quasi)-conserved quantum number $B'$.

\item[2.] dark matter (or its constituents) must have an annihilation cross section much larger
  than electroweak $\sigma_{\rm ann} \gg3~10^{-26}{\rm cm}^3/{\rm sec}$, to ensure that
  $\Omega_{\rm DM}$ is determined dominantly by the  $B'$ asymmetry.
\end{itemize}

The first condition ensures that a global quantum number corresponding to a
linear combination of $B$, $L$ and $B'$ has a weak anomaly, and thus dark matter
carrying $B'$ charge is produced in anomalous processes together with
left-handed quarks and leptons \cite{Kaplan:1991ah,Barr:1990ca}.  At
temperatures $T\gg T_*$ electroweak anomalous processes are in thermal
equilibrium, and equilibrate the various asymmetries $Y_{\Delta B}=c_L
Y_{\Delta L}=c_{B'}Y_{\Delta B'}\sim {\cal O}(10^{-10})$. Here the
$Y_\Delta$'s represent the difference in particle number densities 
normalized to the entropy density $s$, e.g.  $Y_{\Delta B} = (n_B-\bar
n_B)/s$. These are convenient quantities since they are conserved during the
universe thermal evolution.  

At $T\gg M_{\rm DM}$ all  particle masses can be neglected,
and $c_L$ and $c_{B'}$ are order one coefficients,  determined via chemical equilibrium conditions enforced 
by elementary reactions faster
than the universe expansion rate~\cite{Harvey:1990qw}. These coefficients can be computed
in terms of the particle
content,  finding e.g. $c_L=-28/51$ in the standard model and $c_L=-8/15$ in the Minimal Supersymmetric Standard Model.  

At $T\ll M_{\rm DM}$, the $B'$ asymmetry gets suppressed by a Boltzmann
exponential factor $e^{-M_{\rm DM}/T}$.  A key feature of sphaleron
transitions is that their rate gets suddenly suppressed at some temperature
$T_*$ slightly below the critical temperature at which $SU(2)_L$ starts to be
spontaneously broken.  Thereby, if $M_{\rm DM}<T_*$ the $B'$ asymmetry gets
frozen at a value of ${\cal O}(Y_{\Delta B})$, while  if instead $M_{\rm DM} >
T_*$ it gets exponentially suppressed as $Y_{\Delta B'}/Y_{\Delta
  B}\sim e^{-M_{\rm DM}/T}$.
 
More in detail, the sphaleron processes relate the asymmetries of the various
fermionic species with chiral electroweak interactions as follows. If
$B^{\prime}$, $B$ and $L$ are the only quantum numbers involved then the
relation is:
\begin{equation} 
\frac{Y_{\Delta B^{\prime}}}{Y_{\Delta B} } =c\cdot  {\cal S}\left(\frac{M_{\rm DM}}{T_*}\right),\qquad
c=\bar{c}_{B^{\prime}}+ \bar{c}_L \frac{Y_{\Delta L}}{Y_{\Delta B} } \ ,
\end{equation}
where the order-one $\bar{c}_{L,B^{\prime}}$ coefficients are related to the
$c_{L,B^{\prime}}$ above in a simple way.  The explicit numerical values of
these coefficients depend also on the order of the finite temperature
electroweak phase transition via the imposition or not of the weak isospin
charge neutrality. In \cite{Ryttov:2008xe,Gudnason:2006yj} the dependence on
the order of the electroweak phase transition was studied in two explicit
models, and it was found that in all cases the coefficients remain of order
one. The statistical function ${\cal S}$ is:
\begin{equation}
{\cal S} (z) = \left\{ \begin{array}{rl}
\frac{6}{4\pi^2} \int_{0}^{\infty} dx\ x^2\cosh^{-2} \left( \frac{1}{2} \sqrt{x^2 + z^2 } \right) &\qquad  {\rm for~fermions} \ , \\
\frac{6}{4\pi^2} \int_{0}^{\infty} dx\ x^2\sinh^{-2} \left( \frac{1}{2} \sqrt{x^2 + z^2 } \right) &\qquad {\rm for~bosons} \ .
\end{array} \right.
\end{equation}
with $S(0) = 1 (2)$ for bosons (fermions) and $S(z) \simeq 12~(z/2\pi)^{3/2} e^{-z}$ at $z\gg 1$.
We assumed the standard model fields to be relativistic and checked that this is a good approximation even for the top quark \cite{Gudnason:2006yj,Ryttov:2008xe}. 
The statistic function leads to the two limiting results:
\begin{equation}
\frac{Y_{\Delta B'}}{Y_{\Delta B}} =c\times  \left\{
 \begin{array}{cc}
{\cal S}(0)  
&  \qquad  {\rm for } \quad M_{\rm DM}\ll  T_* \\
12\left({M_{\rm DM}}/{2\pi T_*}\right)^{3/2}\,e^{-M_{\rm DM}/T_*} & \qquad {\rm for }\quad M_{\rm DM}\gg T_* 
\end{array}
 \right. \quad.
\end{equation}

Under the assumption that all antiparticles carrying $B$ and $B'$ charges are
annihilated away we have $Y_{\Delta B'}/Y_{\Delta B}=n_{B'}/n_B$. The observed dark matter density
\begin{equation}
\frac{\Omega_{\rm DM}}{\Omega_B}=\frac{M_{\rm DM}\,n_{B'}}{m_{p}\,n_B}\approx 5
\end{equation}
(where $m_p\approx 1\,$GeV) can be reproduced for two possible 
values of the dark matter mass:
\begin{itemize}
\item[i)]  $M_{\rm DM} \sim 5\,$GeV if $M_{\rm DM}\ll T_*$, times model dependent order one coefficients.
\item[ii)] $M_{\rm DM}\sim  8\, T_*\sim 2 ~{\rm TeV}$ if $M_{\rm DM}\gg T_*$,
with a mild dependence on the model-dependent order unity coefficients.
\end{itemize}
The first solution is well known~\cite{Kaplan:1991ah} and corresponds to a light dark matter candidate. While the second condition would lead to a dark matter candidate with a mass of the order of the electroweak scale. This is the asymmetric dark matter paradigm. Many related properties (valid also for symmetric type scenarios) are not yet constrained by our current knowledge of dark matter,
for example the specific dark matter candidate may or may not be a stable particle~\cite{Nardi:2008ix} and
it may or may not be identified with its antiparticle~\cite{Nussinov:1985xr}. However, very recently it was shown that there is a neat way to {\it discover} the existence asymmetric dark matter by studying the cosmic sum rules introduced in \cite{Masina:2011hu,Frandsen:2010mr}.

Asymmetric dark matter candidates were put forward in \cite{Nussinov:1985xr} as
technibaryons, in \cite{Gudnason:2006ug} as Goldstone bosons, and subsequently
in many diverse forms \cite{Foadi:2008qv,Khlopov:2008ty,Dietrich:2006cm,Sannino:2009za,Ryttov:2008xe,Kaplan:2009ag,Frandsen:2009mi}.
There is also the possibility of mixed dark matter~\cite{Belyaev:2010kp}, i.e.\ having
both a thermally-produced symmetric component and an asymmetric one.

From the experimental point of view null results from several experiments, such as CDMS~\cite{Ahmed:2010wy} and
Xenon10/100~\cite{Angle:2011th,Aprile:2011hi},
have placed stringent constraints on WIMP-nucleon cross sections.
Interestingly DAMA~\cite{Bernabei:2008yi} and CoGeNT~\cite{Aalseth:2010vx} have
both produced evidence for an annual modulation signature for dark matter, as
expected due to the relative motion of the Earth with respect to the dark matter halo.
These results support a light WIMP with mass of order a few GeV, which offers
the attractive possibility of a common mechanism for baryogenesis
and dark matter production. At first glance it seems that the WIMP-nucleon cross
sections required by DAMA and CoGeNT have been excluded by CDMS and Xenon upon
assuming spin-independent interactions between WIMPs and nuclei (with protons
and neutrons coupling  similarly to WIMPs), however a number of resolutions for
this puzzle have been proposed in the literature~\cite{Khlopov:2010pq,TuckerSmith:2001hy,Chang:2010yk,Feng:2011vu,Frandsen:2011ts,DelNobile:2011je,DelNobile:2011yb}. Interestingly, also recent results from  the CRESST-II experiment report signals of light dark matter \cite{Angloher:2011uu}. 

A composite origin of dark matter, along the lines detailed above, is therefore quite an intriguing possibility given that the
bright side of the universe, constituted mostly by nucleons, is also composite.
Thus a new strongly-coupled theory could be at the heart of dark matter. Furthermore for the first time on the lattice, a
technicolor-type extension of the standard model, expected to naturally yield a light dark matter candidate, as introduced in \cite{Ryttov:2008xe} and used in
\cite{DelNobile:2011je} to reconcile the experimental observations, has been investigated \cite{Lewis:2011zb}. Here it was shown that strongly interacting theories can, indeed, support electroweak symmetry breaking while yielding natural light dark matter candidates.  

%
 
Models of
dynamical breaking of the electroweak symmetry do support the possibility of 
generating the experimentally observed baryon (and possibly also the technibaryon/dark matter)
asymmetry of the universe directly at the electroweak phase
transition~\cite{Cline:2008hr,Jarvinen:2009wr,Jarvinen:2009pk}.  Electroweak baryogenesis
\cite{Shaposhnikov:1986jp} is, however, impossible in the standard model
\cite{Kajantie:1995kf}.

\subsection{Composite Inflation}
Another prominent physics problem is inflation \cite{Starobinsky:1979ty,Starobinsky:1980te,Mukhanov:1981xt,Guth:1980zm,Linde:1981mu,Albrecht:1982wi}, the mechanism responsible for an early rapid expansion of our universe. Inflation, similar to the standard model Higgs mechanism, is also modeled traditionally via the introduction of new scalar fields. However, field theories featuring fundamental scalars are unnatural. The reason being that typically these theories lead to the introduction of symmetry-unprotected super-renormalizable operators, such as the scalar quadratic mass operator. Quantum corrections, therefore, introduce untamed divergencies which have to be fine-tuned away.  Following the composite Higgs section  \cite{Sannino:2009za,arXiv:0804.0182} one can imagine a new natural strong dynamics underlying the inflationary mechanism  \cite{Channuie:2011rq}.  In \cite{Bezrukov:2011mv} we spelled out the setup for generic models of composite inflation.  Another logical possibility is that theories with scalars are gauge-dual to theories featuring only fermionic degrees of freedom \cite{Sannino:2009qc,Sannino:2010fh,Sannino:2011mr,Mojaza:2011rw}.  

We briefly review here the general setup for strongly coupled inflation  \cite{Channuie:2011rq,Bezrukov:2011mv}. We start by identifying the inflaton with one of the lightest composite states of a generic strongly coupled theory and denote it with $\Phi$. This state has mass dimension $d$. This is the physical dimension coming from the sum of the engineering dimensions of the elementary fields constituting the inflaton augmented by the anomalous dimensions due to quantum corrections in the underlying gauge theory. {}We concentrate  \cite{Channuie:2011rq}  on the non-Goldstone sector of the theory\footnote{The Goldstone sector, if any,  associated to the potential dynamical spontaneous breaking of some global symmetries of the underlying gauge theory will be investigated elsewhere}. 

We then consider the following coupling to gravity in the Jordan frame: 
\begin{eqnarray}
\mathcal{S}_{CI,J}=\int d^{4}x \sqrt{-g}\left[-\frac{{\cal M}^{2}+\xi\,{\Phi}^{\frac{2}{d}}}{2}g^{\mu\nu}R_{\mu\nu}+\mathcal{L}_{\Phi}\right], ~~ \mathcal{L}_{\Phi}=g^{\mu\nu}\Phi^{\frac{2-2d}{d}}\partial_{\mu}\Phi\partial_{\nu}\Phi-V({\Phi}), \label{nonminimal}
\end{eqnarray}
with $\mathcal{L}_{\Phi}$ the low energy effective Lagrangian for the field $\Phi$ constrained by the symmetries of the underlying strongly coupled theory.  In this framework ${\cal M}$ is not automatically the Planck constant $M_{Pl}$. The non-minimal coupling to gravity is controlled by the dimensionless coupling $\xi$. The non-analytic power of $\Phi$  emerges because we are requiring a dimensionless coupling with the Ricci scalar. Abandoning the conformality requirement allows for operators with integer powers of $\Phi$ when coupling to the Ricci scalar. However a new energy scale must be introduced to match the mass dimensions. 

We diagonalize the gravity-composite dynamics model via the conformal transformation: \begin{eqnarray}
g_{\mu\nu}\rightarrow\tilde{g}_{\mu\nu}=\Omega({\Phi})^2 g_{\mu\nu},\quad\Omega({\Phi})^2=\frac{{\cal M}^2+\xi\Phi^{\frac{2}{d}}}{M_{p}^2},
\end{eqnarray}
such that 
\begin{eqnarray}
\quad\tilde{g}^{\mu \nu}=\Omega^{-2}g^{\mu\nu},\quad\sqrt{-\tilde{g}}=\Omega^4\sqrt{-g}.
\end{eqnarray}
 We use both the Palatini and the metric formulation. The difference between the two formulations resides in the fact that in the Palatini formulation the connection $\Gamma$ is assumed not to be directly associated with the metric $g_{\mu\nu}$. Hence the Ricci tensor $R_{\mu\nu}$ does not transform under the conformal transformation.  
 
Applying the conformal transformation we arrive at the Einstein frame action:
\begin{eqnarray}
\mathcal{S}_{CI,E} =\int d^{4}x \sqrt{-g}\left[ -\frac{1}{2} M_{p}^2 \,\, g^{\mu \nu}R_{\mu \nu} + \Omega^{-2} \left(\Phi^{\frac{2-2d}{d}} + f \cdot 3 M_p ^2 {\Omega'} ^{2} \right)g^{\mu \nu} \partial_{\mu} \Phi \partial_{\nu} \Phi  - \Omega ^{-4} V({\Phi}) \right]. \nonumber \\
\end{eqnarray}
Primes denote derivatives with respect to $\Phi$ and tildes are dropped for convenience. $f=1$ signifies the metric formulation \cite{Kaiser:1994vs,Tsujikawa:2000wc,Bezrukov:2008ut,Barvinsky:2008ia} and $f=0$ the Palatini one \cite{Bauer:2010jg}.  

We introduce a canonically normalized field $\chi$ related to $\Phi$ via
\begin{eqnarray}
\frac{1}{2} \tilde{g}^{\mu \nu} \partial_{\mu} \chi (\Phi) \partial_{\nu} \chi(\Phi) = \frac{1}{2} \left( \frac{d \chi}{d \Phi} \right)^2 \tilde{g}^{\mu \nu} \partial_{\mu} \Phi \partial_{\nu} \Phi \ ,
\end{eqnarray}
with
\begin{eqnarray}
\frac{1}{2} \left( \frac{d \chi}{d \Phi} \right)^2 &= \Omega^{-2} \left(\Phi ^{\frac{2-2d}{d}} + f \cdot 3 M_p ^2 {\Omega'} ^2 \right) = \Omega^{-2}\left(1 +  f \cdot\frac{3 \xi ^2}{d^2 M_{p}^2} \Omega^{-2} \Phi^{\frac{2}{d}} \right) \Phi ^{\frac{2-2d}{d}}. \label{defchi}
\end{eqnarray}
In terms of the canonically normalized field we have: 
\begin{eqnarray}
\mathcal{S}_{CI,E} &=\int d^{4}x \sqrt{-g}\left[-\frac{1}{2} M_{p}^2 g^{\mu \nu}R_{\mu \nu} + \frac{1}{2} g^{\mu \nu} \partial_{\mu} \chi \partial_{\nu} \chi- U(\chi)  \right].
\end{eqnarray}
With
\begin{eqnarray}
U(\chi) \equiv \Omega^{-4}V(\Phi).  
\end{eqnarray}
Within this framework we determined in \cite{Bezrukov:2011mv} useful expressions for the slow-roll parameters for composite inflation and provided the explicit example in which the inflaton emerges as the lightest glueball field associated to, in absence of gravity, a pure Yang-Mills theory. This theory constitutes the archetype of any composite model in flat space and consequently of models of composite inflation. We showed  that it is possible to achieve successful glueball inflation. Furthermore the natural scale of compositeness associated to the underlying Yang-Mills gauge theory, for the consistence of the model, turns to be of the order of the grand unified scale. This result is in agreement with the scale of compositeness scale determined in \cite{Channuie:2011rq} for a very different underlying model of composite inflation. 

We also demonstrated that, in the metric formulation, that unitarity-cutoff for inflaton-inflaton scattering is well above the energy scale relevant for composite inflation. It is now possible to envision a large number of new avenues to explore within this class of models. 
  
\section{Conclusion} 
I hope that this concise review is able to convey the message of how new strong dynamics \cite{Sannino:2004qp,Dietrich:2006cm,Ryttov:2007sr,Ryttov:2007cx,Pica:2010mt,Pica:2010xq,Ryttov:2010iz} can be efficiently used to address the relevant challenges in particle physics and cosmology from composite Higgs dynamics to dark matter and inflation.

\acknowledgments
It is a pleasure to thank P. Channuie,  C. Kouvaris, M. Mojaza, C. Pica,  and U. Ish\o j S\o ndergaard for comments and careful reading of the manuscript.

\end{document}